\documentclass{article}

\usepackage[english]{babel}
\usepackage{bbm}
\usepackage[latin1]{inputenc}
\usepackage{amsmath}
\usepackage{amssymb}
\usepackage{natbib}
\usepackage{graphicx}
\usepackage{hyperref}
\usepackage{tikz}
\usepackage{setspace}
\usepackage{lineno}

\newcommand{\pkg}[1]{{\normalfont\fontseries{b}\selectfont #1}}
\let\proglang=\textsf
\let\code=\texttt

\title{Goodness-of-fit statistics for approximate Bayesian computation.}
\author{Louisiane Lemaire$^{1,2}$ \and Flora Jay$^{3,4}$  \and   I-Hung Lee$^{1,2}$ \and Katalin Csill\'ery$^5$ \and Michael G. B. Blum$^{1,2}$}

\begin{document}
\maketitle
\noindent$^{1}$ Univ. Grenoble Alpes, TIMC-IMAG, Grenoble, France\\
\noindent$^{2}$ CNRS, TIMC-IMAG, Grenoble, France\\
\noindent$^{3}$ UMR 7206, Mus\'eum National d'Histoire Naturelle, CNRS, Universit\'e Paris 7 Diderot, Paris France\\
\noindent$^{4}$ Laboratoire de Recherche en Informatique, CNRS UMR 8623, Universit\'e Paris-Sud, Orsay, France\\
\noindent$^{5}$ ETH Z\"urich, Adaptation to a Changing Environment, Institute of Integrative Biology, Universitaetstrasse 16, 8092 Zurich, Switzerland\\

Corresponding author

Michael G.B. Blum

Laboratoire TIMC-IMAG

Facult\'e de m\'edecine

38706 La Tronche, France

michael.blum@imag.fr

\begin{description}
	\item[Running head] Goodness-of-fit with ABC
\end{description}

\clearpage

%
%
%
\section*{Abstract}

\doublespacing
Approximate Bayesian computation is a statistical framework that uses numerical simulations to calibrate and compare models. Instead of computing likelihood functions, Approximate Bayesian computation relies on numerical simulations, which makes it applicable to complex models in ecology and evolution. As usual for statistical modeling, evaluating goodness-of-fit is a fundamental step for Approximate Bayesian Computation. 

Here, we introduce a goodness-of-fit approach based on hypothesis-testing. We introduce two test statistics based on the mean distance between numerical summaries of the data and simulated ones. One test statistic relies on summaries simulated with the prior predictive distribution whereas the other one relies on simulations from the posterior predictive distribution.  For different coalescent models, we find that the statistics are well calibrated, meaning that the type I error can be controlled. However, the statistical power of the two statistics is extremely variable across models ranging from $20\%$ to $100\%$. The difference of power between the two statistics is negligible in models of demographic inference but substantial in an additional and purely statistical example. When analyzing resequencing data to evaluate models of human demography, the two statistics provide similar results and confirm that an out-of-Africa bottleneck cannot be rejected for Asiatic and European data. We also consider two speciation models in the context of a butterfly species complex. One goodness-of-fit statistic indicates a poor fit for both models, and the numerical summaries causing the poor fit were identified using posterior predictive checks.

Statistical tests for goodness-of-fit should foster evaluation of model fit in Approximate Bayesian Computation. The test statistic based on simulations from the prior predictive distribution is implemented in the {\code gfit} function of the \textbf{\textsf{R}} {\it abc} package.
%
%
%
%
%
%
%
%
%
%
%
%
%
%
%
%


\doublespacing

\clearpage

\section*{Introduction} 
Evaluating the goodness-of-fit of a statistical model is part of statistical modeling. Evaluating to what extent a model fit the data is a prerequisite before model improvement, which is the third step of Bayesian data analysis following model formulation and model fitting \cite[]{gelman14}. Approximate Bayesian computation (ABC) follows the rules of Bayesian data analysis and should also encompass goodness-of-fit evaluation \cite[]{csilleryetal10}.

Ecological or evolutionary models fitted and compared with ABC are usually introduced for explanatory purposes. The objective is to explain the data in terms of ecological and evolutionary processes that arose in the past. Typical questions addressed with ABC are related to historical processes of speciation \cite[]{roux13}, processes of divergence and migration between populations \cite[]{laval10,pelletier14}, processes of biological adaptation \cite[]{peter12}, or ecological dynamics of natural ecosystems \cite[]{hartig14,lagarrigues15}. There is another use of statistical modeling in ecology that seeks a predictive goal instead of an explanatory one. Species distribution models are examples of statistical models introduced for sake of prediction  \cite[]{elith09}. Models designed for a predictive purpose can be evaluated using cross-validation techniques to measure predictive accuracy \cite[]{hijmans12}. However, there are no measures of predictive accuracy for explanatory models. The impossibility to evaluate prediction ability makes goodness-of-fit evaluation all the more important for models introduced for explanatory purposes.


%
%
%
%
%
%
%
%

Evaluating goodness-of-fit in Bayesian analysis is usually performed with graphical checks such as posterior predictive checks \cite[]{gelman14,gruenstaeudl15}. It consists in simulating the parameter $\theta$ according to the posterior distribution $p(\theta|s)$, where $s$ denotes the observed summary statistics computed from the data, and then to generate replicated summary statistics $s^{\rm rep}$ based on the generating mechanism $p(s|\theta)$. Summary statistics simulated with this mechanism are sampled according to the {\it posterior predictive distribution} that is denoted $p(s^{rep}|s)$ \cite[]{gelman14}. Observed summary statistics $s$ are then compared to the one-dimensional histograms of these replicated summary statistics. Finding an observed  summary statistic outside of the range of the posterior predictive distribution is an indication of poor fit. It is also possible to compute the fraction of times posterior predictive simulations are larger (or lower) than the observed summary statistic to obtain posterior predictive P-values \cite[]{meng94}. Posterior predictive checks are well-suited to Approximate Bayesian Computation for at least two reasons. First, parameter inference is based on summary statistics $s$, which provide straightforward test statistics for posterior predictive checks. Second, the simulation mechanism, which consists of simulating statistics according to $p(s|\theta)$, is already used for parameter inference and can be recycled for goodness-of-fit. Applications of ABC in ecology and evolution have used posterior predictive checks to evaluate model fit in different fields such as demographic inferences in population genetics \cite[]{li14}, taxonomy and systematics \cite[]{dong14}, or ecosystem modeling \cite[]{morales15}.

%
%


However, one major concern about posterior predictive P-values is that they are not properly calibrated. Posterior predictive P-values are not uniformly distributed when the data are realizations of the investigated model (i.e. when the null hypothesis is true). Posterior predictive P-values are more concentrated around 1/2 than expected under a uniform distribution \cite[]{robins00}. To provide well-calibrated P-values, we introduce an alternative approach for performing goodness-of-fit in ABC. The objective of the proposed goodness-of-fit (GOF) statistics is to provide an assessment of model fit based on a classical hypothesis testing framework, where each investigated model serves as the null hypothesis. Providing well-calibrated P-values allows for its common interpretation across statistical problems in ecology. In Bayesian statistics, there have been already several attempts at providing well-calibrated P-values, including conditional predictive P-values or partial posterior predictive P-values but they are difficult to compute in complex statistical models \cite[]{bayarri00}. Another proposition includes a Bayesian chi-square statistic for GOF but it is limited to uni-dimensional data or summary statistic \cite[]{johnson04}. To provide well-calibrated P-values, we introduce two GOF test statistics whose computations are straightforward with ABC algorithms. P-values are evaluated based on the histogram that is constructed by repeatedly computing the GOF statistics on pseudo-observed data. After providing definitions of the two GOF statistics, we evaluate their statistical properties in different models of interest in population genetics, and we compute the statistics in the context of human and butterfly molecular data.

\section*{Methods}
%
%
%

\subsection*{A goodness-of-fit statistic based on the prior predictive distribution}

\begin{figure}
\centering
   \includegraphics[width=15cm]{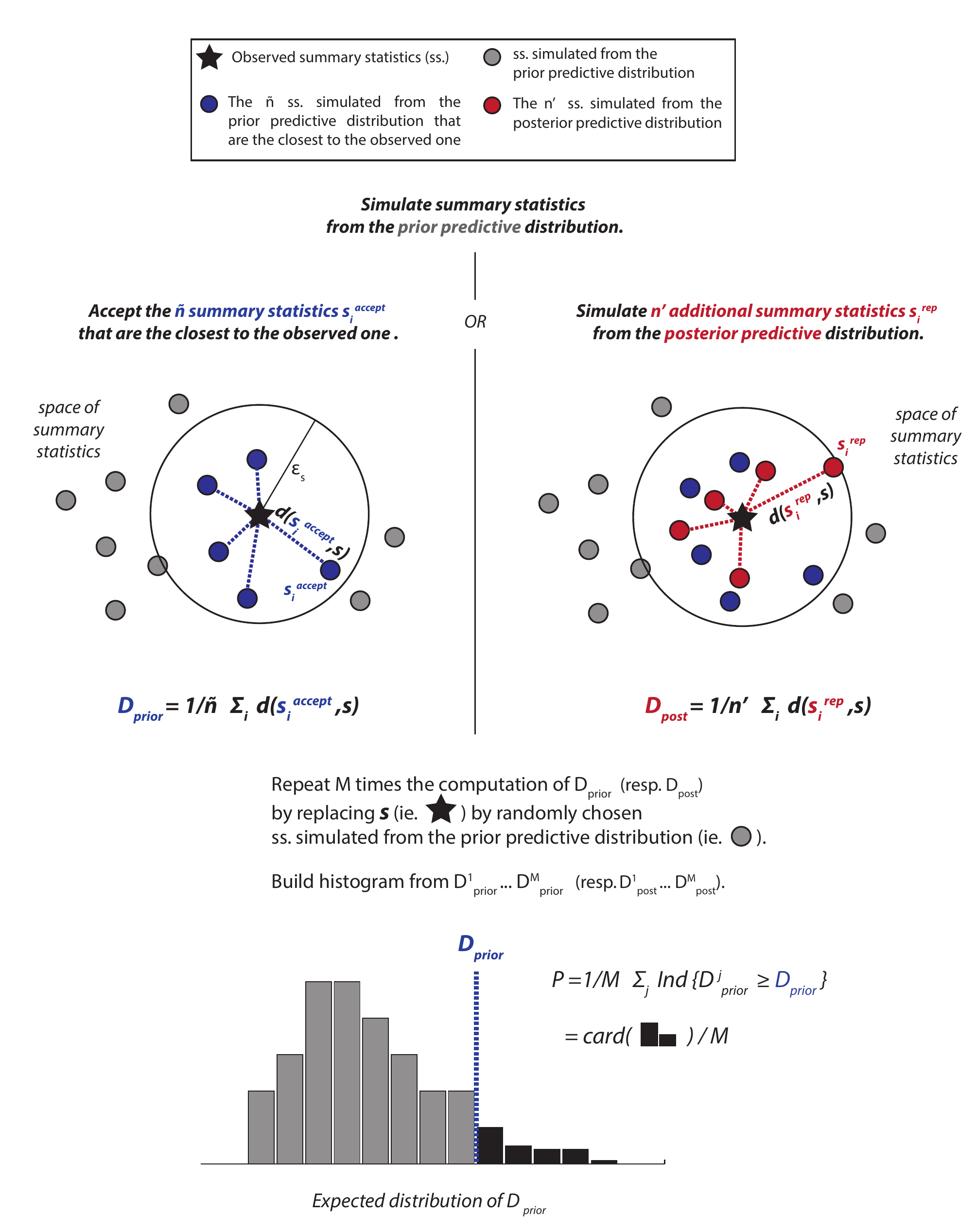}
       \caption{Graphical description of the two goodness-of-fit statistics. The number of accepted summary statistics is denoted as $\tilde{n}$, the observed summary statistic is denoted as $s$, the accepted summary statistics are denoted as $s_1^{\rm accept},\dots,s_{\tilde{n}}^{\rm accept}$, the summary statistics simulated with the posterior predictive distribution are denoted as $s_1^{\rm rep},\dots,s_{n'}^{\rm rep}$ , the number of posterior replicates is denoted as $n'$, the number of pseudo-observed datasets used to compute P-values is denoted as $M$, and the indicator function is denoted as ${\rm Ind}$.}
    \label{fig:flora}
\end{figure}

The objective is to test a null hypothesis that assumes that the data are realizations of a statistical model denoted $\mathcal{M}_0$. The statistical model $\mathcal{M}_0$ is defined by a possibly multivariate parameter denoted $\theta$. In order to introduce the goodness-of-fit statistic, we recall what is the rejection algorithm. The rejection algorithm is the basic algorithm to produce samples from a distribution that approximates the posterior distribution of $\theta$ \cite[]{pritchard99}. First, parameter values, $\theta_i$,  $i=1,\dots, n$, are sampled from the prior distribution $p(\theta)$. Then, summary statistics, $s_i$,  are simulated using the generating mechanism $p(s | \theta_i)$. The resulting distribution with which summary statistics are simulated is named as the {\it prior predictive distribution} \cite[]{gelman14}. Simulated summary statistics are compared to the observed ones using a distance measure $d$, such as the Euclidean distance. The rejection algorithm rejects all simulations that are too far from the simulations based on the distance measure $d$. In practice, the percentage $\tilde{n}/n$ of accepted simulations, coined as the acceptance rate, is set to a given value (e.g. $1\%$). The goodness-of-fit statistic is defined as the mean distance between the observed summary statistics and the $\tilde{n}$ simulated statistics that have been accepted (Figure \ref{fig:flora})

%
%
\begin{equation}
\label{eq:gof}
D_{\rm prior}=\frac{1}{\tilde{n}}\sum_{i=1}^{\tilde{n}} d(s_i^{\rm accept},s),
\end{equation}
where  $s_1^{\rm accept},\dots,s_{\tilde{n}}^{\rm accept}$ denote the $\tilde{n}$ accepted simulated summary statistics. To compute the distance $d$, we assume that each one-dimensional summary statistic has been standardized using the median absolute deviation, which is a robust estimate of the standard deviation \cite[]{csillery12}. The median absolute deviation is computed based on summary statistics simulated from the prior predictive distribution.

To obtain the null distribution of the test statistic $D_{\rm prior}$, we consider pseudo-observed data sets \cite[]{bertorellel10}. A simulation from model $\mathcal{M}_0$ is discarded and considered as the observed data. The remaining $n-1$ simulations are then used to perform the rejection algorithm and to compute the test statistic. Repeating this process $M$ times, we obtain a vector of test statistics $D_{\rm prior}^1,\dots,D_{\rm prior}^M$. The P-value $P$ of the goodness-of-fit procedure is computed as the proportion of test statistics obtained under model $\mathcal{M}_0$ that are larger than the observed one
%
%

\begin{equation}
\label{eq:pval}
P=\frac{1}{M} \sum_{j=1}^M \mathbbm{1}_{D_{\rm prior}^j \geq D_{\rm prior}},
\end{equation}
where $\mathbbm{1}$ denotes the indicator function. By construction, P-values will be uniformly distributed for summary statistics simulated under the prior predictive distribution, which is defined by the prior distribution of the parameters $p(\theta)$ and the generating mechanism for the summary statistics $p(s | \theta)$.

%
%
%
%
%
Because computing the goodness-of-fit statistic of equation (\ref{eq:gof}) and its corresponding P-value does not require new simulations in addition to the ones performed for parameter inference, it was possible to implement it in the {\it abc} \textbf{\textsf{R}} package, and the name of the \textbf{\textsf{R}} function is {\code gfit} \cite[]{csillery12}.
%
%

\subsection*{An alternative goodness-of-fit statistic based on the posterior predictive distribution}

%
%

We derive an alternative statistic based on summary statistics simulated with the posterior predictive distribution. The alternative statistic denoted as $D_{\rm post}$ measures the mean distance between observed summary statistics and statistics simulated based on parameters sampled from the posterior distribution. An advantage of $D_{\rm post}$ is that it can make use of regression-adjustments that improve the estimation of the posterior distribution \cite[]{beaumont02,blumfrancois10}. The statistic $D_{\rm post}$ is defined as follows
%
%
%
%
%
%
%
%
\begin{equation}
\label{eq:gof_post}
D_{\rm post}=\frac{1}{n'}\sum_{i=1}^{n'} d(s^{\rm rep}_i,s),
\end{equation}
where $n'$ denotes the number of posterior replicates, and where the summary statistics $s^{\rm rep}_i$, $i=1,\dots,n'$, have been sampled according to the posterior predictive distribution $p(s^{\rm rep}|s)$. To obtain the null distribution, the test statistic $D_{\rm post}$ is computed for $M$ pseudo-observed data sets, and P-values are obtained similarly to equation (\ref{eq:pval}). The computation of the null distribution is computationally intensive. Computing one P-value requires $M\times n'$ times call to the generating mechanism $p(s|\theta)$ that returns a set of summary statistics based on an input value of the parameter $\theta$. Again, P-values will be uniformly distributed for summary statistics simulated under the prior predictive distribution.






\subsection*{Examples}
\subsubsection*{An example of statistical model}

To evaluate type I error and statistical power, we start by considering a toy statistical model. The objective is to test the goodness-of-fit of a Gaussian distribution and of a Laplace distribution when data were simulated with one of these two possible distributions \cite[]{francois11}. For each possible distribution, we simulated $10,000$ samples, each of them consisting of a sample of size $50$ or $100$ summarized by its mean, variance, skewness and kurtosis. For the Gaussian samples, we consider a uniform prior between $-10$ and $10$ for the mean parameter and  an inverse chi square parameter with 3 degrees of freedom for the variance parameter. For Laplace samples, we consider the same prior for the location parameter. We simulated the scale parameter so that the theoretical variance is also an inverse chi square parameter with 3 degrees of freedom.  Detailed aspects of the simulations can be found in the \textbf{\textsf{R}} file that contains a script to generate the simulations and evaluate type I and II errors (Supplementary file 1).

\subsubsection*{Examples of demographic inference}
\begin{figure}
\centering
   \includegraphics[width=15cm]{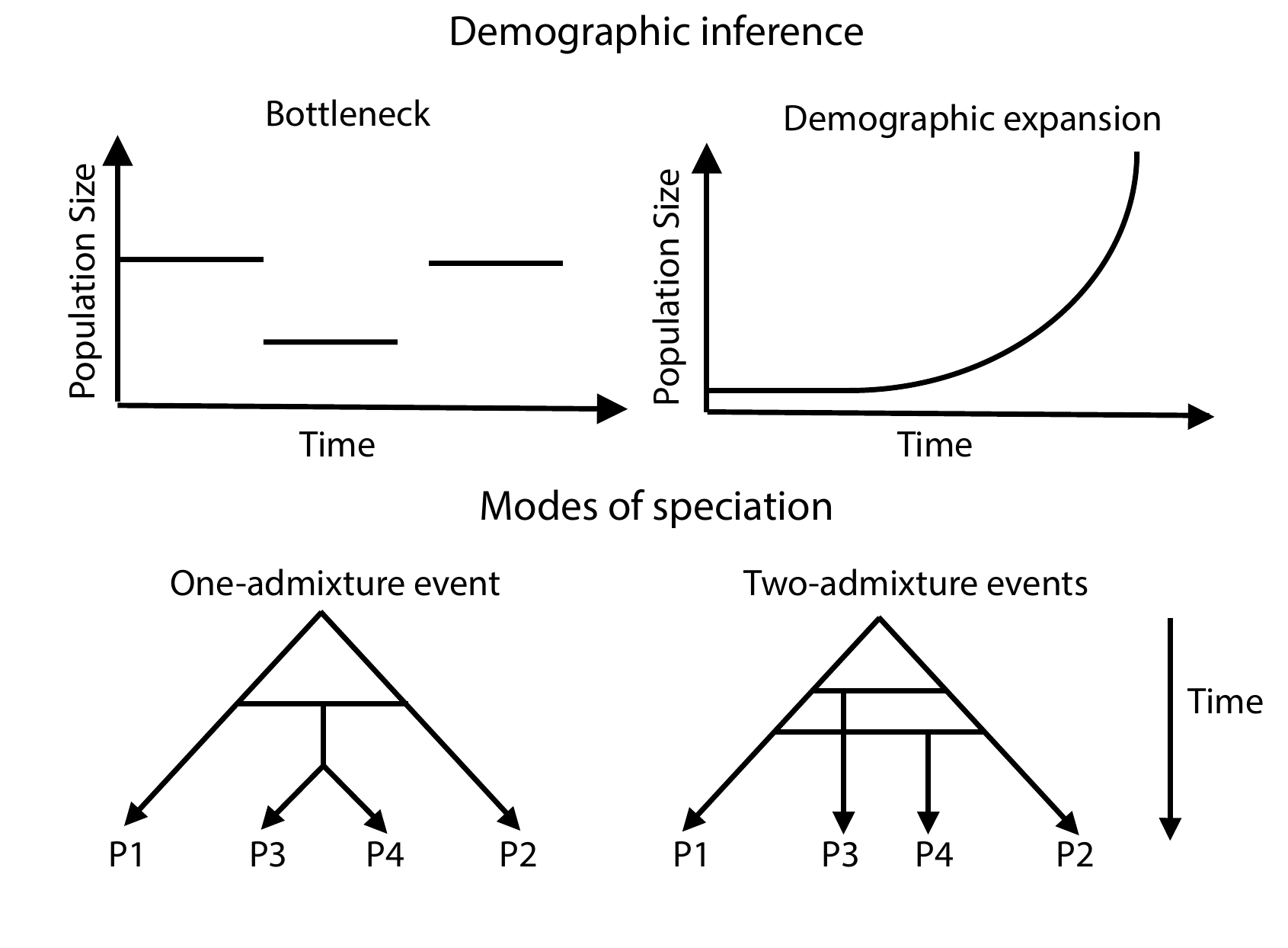}
    \caption{Graphical description of the two examples used to evaluate type I error and power with the goodness-of-fit statistics.}
    \label{fig:models}
\end{figure}

Then, we consider two biologically relevant problems of statistical inference, one related to demographic inference and another to model speciation processes. In the first problem, we test if population genetic data are compatible with a model of historical bottleneck or of population expansion (Figure \ref{fig:models}). For this problem, we consider data simulated  with two different coalescent models for which we also used different sets of summary statistics. 

The first set of simulation\textbf{s} was performed with {\it ms} \cite[]{hudson02} and consists of 50 $2,000$ bp sequences that have been sequenced from 10 diploid individuals. Prior distributions and further details can be found in the \textbf{\textsf{R}} file that contains a script to generate the simulations (Supplementary file 2). A total of $60,000$ simulations was performed for each demographic model. Data were summarized using three summary statistics: average nucleotide diversity, and the mean and variance (over loci) of Tajima's D \cite[]{voight05}. Goodness-of-fit was evaluated based on $D_{\rm prior}$ and $D_{\rm post}$.
%
%
%
%

The second set of simulation\textbf{s} was generated using {\it fastsimcoal2} \cite[]{excoffier13} and consists of a total of 100 independent stretches of the genomes for 10 diploid individuals. Prior distributions are given in the supplementary text. Data were summarized with the total number of SNPs and the unfolded site-frequency-spectrum, defined as the vector counting the number of mutations carried by $i$ chromosomes, for $i$ ranging from $1$ to $19$.  Because this simulation framework is computationally intensive, we evaluated its fit only based on the $D_{\rm prior}$ statistic.

\subsubsection*{Examples of speciation models}
In the second problem, we consider two models of divergence and admixture that correspond to hypothesized scenarios of speciation in the butterfly species complex {\it Coenonympha} \cite[]{capblancq15}. Two species from this complex, {\it C. arcania} and {\it C. gardetta}, are assumed to have diverged 1.5 to 4 millions years ago \cite[]{kodandara09}, while a third species {\it C. darwiniana} is assumed to be the result of admixture between the two ancestral populations \cite[]{capblancq15}. Based on samples from four populations (one population of {\it C. arcania}, one of {\it C. gardetta}, and two populations of {\it C. darwiniana} sampled in France and Switzerland), we test the fit of two alternative models. The first model assumes that the same admixture event is at the origin of two populations of {\it C. darwiniana}, and the second model assumes independent admixture events (Figure \ref{fig:models}). The prior distributions are given in the supplementary text. We computed with {\it DIYABC} a total of 16 summary statistics corresponding to the genetic diversity in each population and to the pairwise  $F_{st}$ and Nei's distances between populations \cite[]{cornuetetal10}. We generated $1,000,000$ simulations for each admixture model. Because this simulation framework is computationally intensive, we evaluated its fit only based on the $D_{\rm prior}$ statistic.
%
%
%

\subsubsection*{Application to real data}
We applied the goodness-of-fit statistics $D_{\rm prior}$ and $D_{\rm post}$ to human data consisting of $50$ $2,000$ bp sequence data sampled in 10 individuals coming from three different populations: Africa (Hausa), Asia (Chinese), and Europe (Italian) \cite[]{voight05}. The summary statistics are the average nucleotide diversity and the mean and variance (over loci) of Tajima's D for three human samples coming from Africa (Hausa), Asia (Chinese), and Europe (Italian). The simulations used to evaluate goodness-of-fit are the simulations of bottleneck and of expansion performed with {\it ms}.
%
%
%

We also applied the goodness-of-statistics $D_{\rm prior}$ to the dataset of SNPs sampled in the four butterfly species. A total of 139 individuals were genotyped, including 33 individuals of {\it C. arcania}, 52 individuals of {\it C. gardetta}, 35 individuals of Swiss {\it C. Darwiniana}, and 19 individuals of French {\it C. darwiniana}  \cite[]{capblancq15}. The final data set used to compute the 16 summary statistics contains 510 polymorphic loci.

\subsubsection*{Parameter settings of the GOF statistics}
When computing the GOF statistic $D_{\rm prior}$ of equation (\ref{eq:gof}), the percentage of accepted simulations for the rejection algorithm was set to $\tilde{n}/n=1\%$. P-values were computed using equation (\ref{eq:pval}) with a total of $M=1,000$ replicates. When computing the GOF statistic $D_{\rm post}$ of equation (\ref{eq:gof_post}), we again consider the Euclidean distance for $d$ and assume that summary statistics have been scaled using median absolute deviations estimated based on the  $M\times n'$ posterior replicates. To evaluate $D_{\rm post}$, the percentage of accepted simulations is set to $1\%$, a linear adjustment is used for parameter inference \cite[]{beaumont02}, and the number $n'$ of posterior replicates is of 100.  Each P-value is evaluated using a total of $M=200$ replicates.
%
%
%

%
%

\section*{Results}

\subsection*{Simulation study of a toy model}

We set the expected type I error at $5\%$ by rejecting the null model when P-values are smaller than $5\%$. For the toy model, tests based on $D_{\rm prior}$ or $D_{\rm post}$ are well calibrated with type I error ranging from $4\%$ to $6\%$ when the nominal type I error is of $5\%$. However, the power to reject the null is very weak for the statistic $D_{\rm prior}$ of equation (\ref{eq:gof}). When rejecting the Gaussian distribution, the power is of $9.5\%$ and when rejecting the Laplace model, the power is of $6.5\%$. When considering the alternative statistic $D_{\rm post}$ based on the posterior predictive distribution, the power is of $51.5\%$ when rejecting the Gaussian distribution and of $5\%$ when rejecting the Laplace distribution. Increasing the sample size from $n=50$ to $n=100$ shows again that the power to reject the Gaussian distribution is increased when using $D_{\rm post}$ (power of $78\%$) instead of $D_{\rm prior}$ (power of $11\%$).

\subsection*{Distributions of P-values for different evolutionary models}

\begin{figure}
\centering
   \includegraphics[width=15cm]{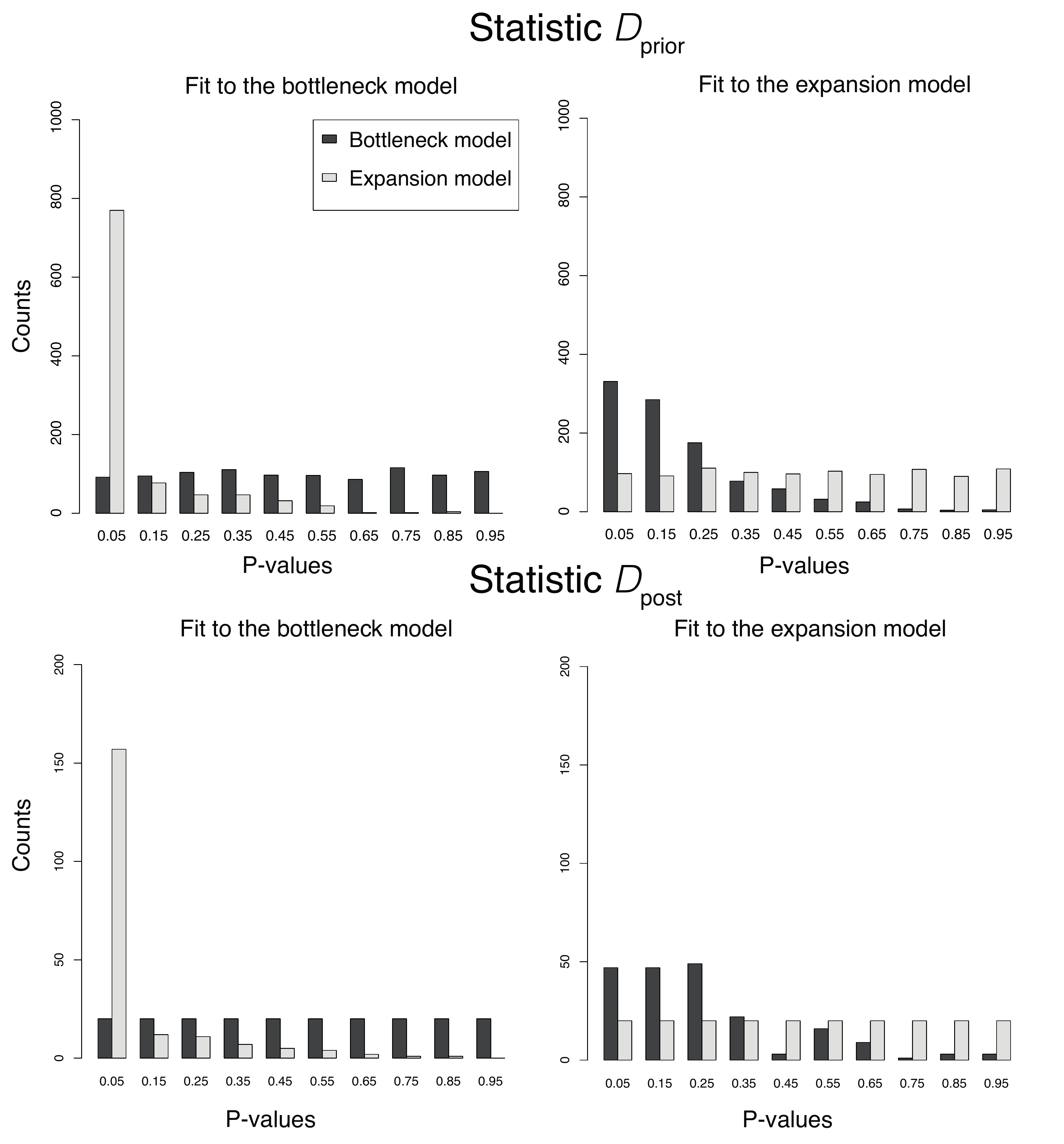}
    \caption{Distribution of P-values for a bottleneck and an expansion model. Data are summarized using three summary statistics, which are the average nucleotide diversity and the mean and variance (over loci) of Tajima's D. When using $D_{\rm prior}$ (resp. $D_{\rm post}$), a total of 1,000 (resp. 200) P-values are computed for each combination of null model and of model used for the simulations.}
  \label{fig:pval_set1}
\end{figure}

Using the first set of simulations for demographic inference, there are two possible null models (bottleneck or expansion) and two possible models for simulating the data (bottleneck or expansion), which results in four different distributions of P-values. When the null model is used for the simulations, the P-values are uniformly distributed (Figure \ref{fig:pval_set1}). When the two models are different, the distributions of P-values are shifted towards zero. Moreover, there is a clear asymmetry when performing model fit. P-values obtained when testing the bottleneck model for simulations of expansion are more shifted towards zero than when testing the expansion model for simulations of bottleneck (Figure \ref{fig:pval_set1}). The distributions of P-values are similar when considering the statistic $D_{\rm post}$ instead of $D_{\rm prior}$ (Figure \ref{fig:pval_set1}).

\begin{figure}
\centering
   \includegraphics[width=15cm]{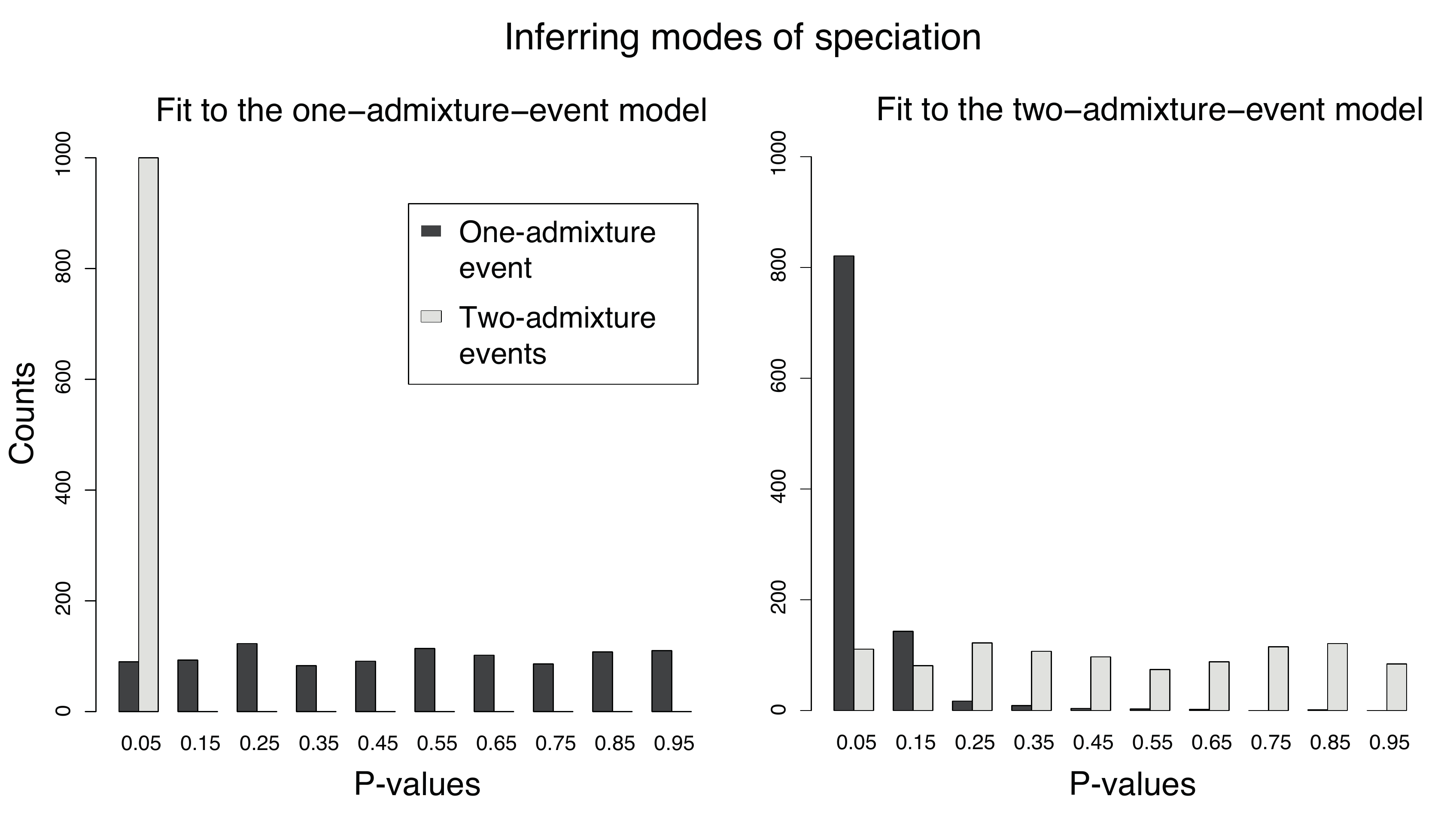}
    \caption{Distribution of P-values based on $D_{\rm prior}$ for models with one and two events of admixture. Data are summarized using 16 summary statistics corresponding to the genetic diversity in each population and to the pairwise $F_{st}$ and Nei distances between pairs of populations. P-values are computed for each combination of null model and of model used for the simulations.}
  \label{fig:pval_butterfly}
\end{figure}

Using the second set of simulations based on the site-frequency-spectrum (SFS) and the simulations of speciation models, we again find that P-values are uniformly distributed when simulations are performed using the null model (Supp Info Figure 1 and Figure \ref{fig:pval_butterfly}). For the SFS based simulations, there is also an asymmetry in the distributions of P-values. P-values obtained when testing the bottleneck model for simulations of expansion are more shifted towards zero than when testing the expansion model for simulations of bottleneck (Supp Info Figure 1). For the speciation models and when the null model is different from the simulation model, P-values obtained when testing the model with one event of admixture are more shifted towards zero than when testing the model with two events of admixture (Figure \ref{fig:pval_butterfly}).

\subsection*{Statistical power}

The power of the test statistic $D_{\rm prior}$ is asymmetric (Table \ref{tab:power}). It is more difficult to reject an expansion model (power of $18\%$ or  $21.5\%$ depending on the summary statistics) than to reject a bottleneck model (power of $67\%$ or of $53\%$ depending on the summary statistics). Finding asymmetric statistical power is expected because P-values obtained when testing the bottleneck model for simulations of expansion  are more shifted toward 0 than when testing the opposite (Figure \ref{fig:pval_set1}). By contrast to the toy example, considering the $D_{\rm post}$ statistic instead of $D_{\rm prior}$ hardly changes statistical power (Table \ref{tab:power}). For speciation models, we also find  asymmetric statistical power. The power is of $99\%$ when rejecting the one-event admixture model whereas it is of $19.5\%$ when rejecting the two-event admixture model. Again, observing asymmetric power is expected because of the shapes of the distributions of P-values (Figure \ref{fig:pval_butterfly}).
%
%



\subsection*{Application to human data}

We apply the goodness-of-statistics $D_{\rm prior}$ and $D_{\rm post}$ to the human resequencing data (Table \ref{tab:data_gof}). The African dataset is compatible with a constant-population size ($P=0.21$), a bottleneck ($P=0.17$), and an expansion model ($P=0.55$). The Asiatic dataset is compatible with a constant-population size and a bottleneck model ($P=0.10$ and $P=0.86$), but not with an expansion model ($P=0.01$). Finally, the European dataset is compatible with a bottleneck model ($P= 0.60$), but both the constant-population size and the expansion models can be rejected ($P=0.02$ and $P<10^{-2}$). Using the goodness-of-fit statistics $D_{\rm post}$ based on posterior replicates leads to similar conclusions (Table \ref{tab:data_gof}). Analysis of the resequencing data with the goodness-of-fit statistics confirms that the out-of-Africa bottleneck cannot be rejected for the Asiatic and the European data \cite[]{voight05}.
%
%
%

\subsection*{Application to study models of speciation in a butterfly species complex}
\begin{figure}
\centering
   \includegraphics[width=15cm]{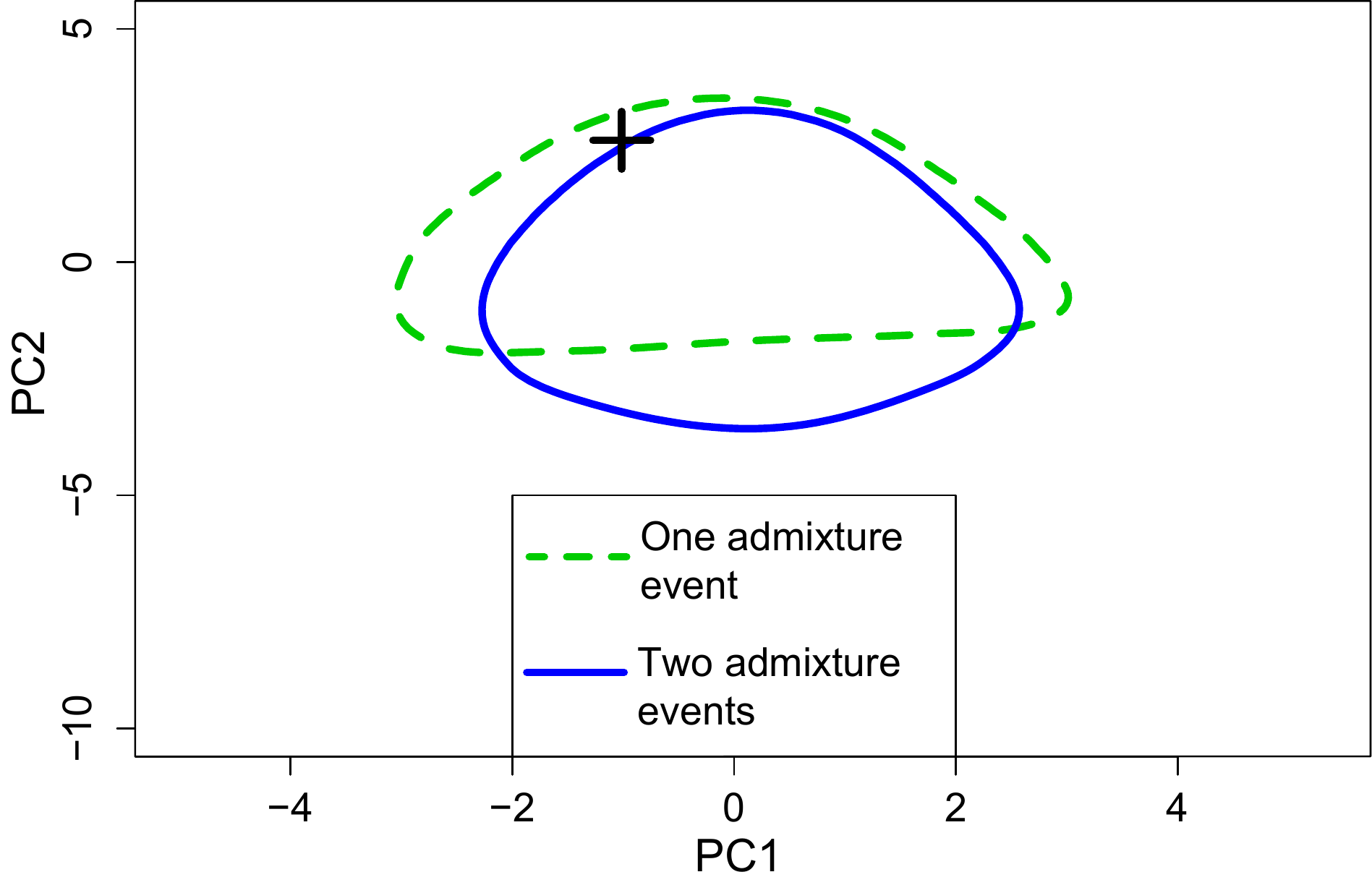}
    \caption{Visual evaluation of the goodness-of-fit of the two admixture scenarios. Principal component analysis applied to the 16 summary statistics simulated based on the prior predictive distribution. The cross corresponds to observed data. The envelopes are obtained using the {\code gfitpca} routine of the {\it abc} package and contain $90\%$ of the simulated data points.}
  \label{fig:pca}
\end{figure}

We fist consider a visualization routine to investigate model fit with ABC. For the two competing models, we computed principal component analysis (PCA) based on the set of 1 million simulations performed under each model and we displayed the  envelope  containing $90\%$ of the points in the space defined by the first two PCs \cite[]{cornuetetal10,sjodin12}. Figure \ref{fig:pca} shows that the model with one admixture event is able to reproduce the summary statistics when summarized by the first two PCs, since the projection of the observed data falls into the $90\%$ envelope. For the two-event admixture model, observed data are outside but nearby the $90\%$ envelope. As a conclusion, the visualization routine based on PCA does not indicate a poor fit of any of the two different admixture models to the polymorphism data of the butterfly {\it Coenonympha} species complex.
%
%
%
%
%
%
%
%
%
%
%
%
%
%
%
%

By contrast, test of GOF based on $D_{\rm prior}$ shows that both models do not provide a good fit to the 16 summary statistics ($P<10^{-4}$). Performing posterior predictive checks confirms the poor fit because 3 out of 16 summary statistics can not be reproduced with any of the two admixture models. In 3 out of 4 populations, the simulated mean genetic diversities across all loci are always smaller than the observed values. Thus, we replace in each population the mean genetic diversity across all loci by the mean genetic diversity across variable loci. After replacement of these 4 summary statistics, the $D_{\rm prior}$ statistic does not indicate a poor fit anymore and both speciation models are able to reproduce the 16 observed summary statistics ($P=0.25$ and $P=0.28$).
%
%
%
%
%
%
%
%
%
%
%
%

\section*{Discussion}

We propose two goodness-of-fit statistics to evaluate model fit in Approximate Bayesian Computation. The first goodness-of-fit statistic $D_{\rm prior}$ is equal to the mean distance between observed summary statistics and the closest simulated ones where summary statistics were simulated with the prior predictive distribution. P-values are estimated based on Monte Carlo replicates by  repeatedly computing the goodness-of-fit statistic on pseudo-observed data or summary statistics simulated from the null model. Based on simulations, we confirm that the statistic is well-calibrated because P-values are uniformly distributed when parameters are simulated with the prior distribution and when summary statistic are simulated with the investigated model. However, its statistical power is extremely variable ranging from $20\%$ to $100\%$ in the simulations we investigated.

Observing variable statistical power is expected for admixture models. For instance, the model that comprises of two admixture events is more flexible than the evolutionary model with one admixture event only. As a consequence, it is more difficult to reject the two-admixture model (power of  $19.5\%$) than the one-admixture model (power of  $99\%$) because most of the simulations obtained with one admixture event can be reproduced with two admixture events but the reverse is not true.
%
%
%
%

%
%
%


We propose a second goodness-of-fit statistic $D_{\rm post}$ based on posterior replicates. It has several advantages. A conceptual advantage is that it is less dependent on the prior that the $D_{\rm prior}$ statistic. When using $D_{\rm prior}$, a good model or evolutionary scenario can come under suspicion with a poor choice of prior distribution. Same arguments were advanced to criticize prior predictive P-values \cite[]{bayarri00}. Another difference between the two goodness-of-fit statistics concerns statistical power. When using uninformative prior distributions, as for the toy statistical model we considered, the $D_{\rm prior}$ statistic has a weak statistical power compared to  $D_{\rm post}$. However, the statistic $D_{\rm post}$ has an important drawback related to its computational burden. When using for instance $n'=100$  posterior replicates to evaluate the statistic and  $M=1,000$ pseudo observed data to compute the null distribution, a total of $100,000=1,000*100$  additional simulations are required to evaluate P-values whereas the $D_{\rm prior}$ statistic does not require  simulations in addition to the ones performed for parameter inference. Simulation-based comparisons between $D_{\rm prior}$ and $D_{\rm post}$ are equivocal. For the simple toy model, the statistical power of $D_{\rm post}$ is $5-7$ times larger than $D_{\rm prior}$ whereas there was no substantial difference for the examples of evolutionary scenarios. 
%
%
%
%
%

Compared to model comparison, which is a common step in ABC, goodness-of-fit has been too neglected. Goodness-of-fit and model comparisons are two different aspects of statistical analysis that should not be confused. Goodness-of-fit tests the absolute fit of a statistical model and do not seek to compare models. The fact that goodness-of-fit provides an absolute measure is valid when considering $D_{\rm prior}$ and $D_{\rm post}$ because they rely on Fisher's approach of significance testing, which requires only one hypothesis, and not on  Neyman-Pearson's approach of hypothesis testing (e.g. likelihood-ratio test) that would require two hypotheses \cite[]{lehmann93,beaumontetal10}. By contrast, model selection provides statistical measures that are relative to the set of models to be compared \cite[]{hickerson14,pelletier14}. Model selection does not test models and do not evaluate model fit. If all candidate models provide a poor fit, model selection will not provide any statistical warning, and that should be a strong endeavor to evaluate model fit.

How should we evaluate  model fit based on test statistics and corresponding P-values? Using a standard P-value cutoff of $0.05$, there are two options  whether or not the P-value is smaller than the cutoff value. The first option is when the null model cannot be rejected ($P>0.05$). If the model passes the test, it does not either certify the `truth' of a current scientific theory. \citet{rykiel96} coins the validation procedure as {\it operational validation}, which is defined as a test protocol to check that the model is an adequate representation of the system. However, he stresses that an adequate representation is not a `guarantee that the scientific basis of the model and its internal structure correspond to the actual processes or cause-effect relationships operating in the real system'. Following Popperian philosophy, a P-value larger than $0.05$ is not an indication that the tested model is true but that it can not be rejected. A null model may not be rejected for many reasons including lack of power, which can be due to a poor choice of summary statistics or a small sample size. 
%
%
%

When the proposed model does not pass the test ($P<0.05$), poorly fitted summary statistics can be identified using prior or posterior predictive checks. For instance, with the genetic markers of the butterfly species complex, we identified that mean genetic diversity over all loci including SNPs that do not vary within the population is responsible for the poor fit of the investigated models. This suggests potential model improvements such as including gene flow following admixture, which would increase the mean genetic diversity over all loci by decreasing the number of private SNPs. The example of the speciation models shows that having a single and well-calibrated P-value, rather than using graphical routines, such as the PCA-based graphical routine of Figure \ref{fig:pca},  offers the opportunity for a convenient evaluation of fit. Providing a single well-calibrated P-value for each model is especially useful when there are many summary statistics as it can occur when reconstructing historical demography with molecular data \cite[]{robinson14}. Posterior predictive checks are useful as a second step to detect the summary statistics that explain a poor fit.
%
%
%
%
%
%


The proposed goodness-of-fit statistics seek to foster evaluation of model fit in ABC. However, the proposed test statistics should not encourage black-box analyses with ABC where decision to reject or not a model relies exclusively on the returned P-value. We view the goodness-of test statistics and corresponding P-values as useful diagnostic devices, particularly when screening models with many summary statistics. P-values are one of many ways to quickly alert oneself to some of
the important features of a data set \cite[]{hill96}. To encourage goodness-of-fit evaluation, the \pkg{abc}  \proglang{R}
 package includes the {\code gfit} function that evaluates statistical significance based on the $D_{\rm prior}$ statistic of equation (\ref{eq:gof}). 

\section*{Acknowledgments}
This work has been supported by the LabEx PERSYVAL-Lab (ANR-11-LABX-0025-01). FJ was supported by the ANR Demochips project (ANR-12-BSV7-0012). We acknowledge Thibaut Capblancq for providing the simulations corresponding to the divergence models, and Fr\'ed\'eric Austerlitz for helpful discussions. Part of the demographic simulations were performed on the GenoToul Bioinformatics hardware infrastructure.
%
%

\clearpage

\bibliographystyle{evolution}
\bibliography{lemaireetal}

\begin{thebibliography}{37}
\providecommand{\natexlab}[1]{#1}
\providecommand{\url}[1]{\texttt{#1}}
\providecommand{\urlprefix}{URL }

\bibitem[{Bayarri and Berger(2000)}]{bayarri00}
Bayarri, M. and J.~O. Berger, 2000.
\newblock P values for composite null models.
\newblock Journal of the American Statistical Association 95:1127--1142.

\bibitem[{Beaumont et~al.(2010)Beaumont, Nielsen, Robert, Hey, Gaggiotti,
  Knowles, Estoup, Panchal, Corander, Hickerson et~al.}]{beaumontetal10}
Beaumont, M.~A., R.~Nielsen, C.~Robert, J.~Hey, O.~Gaggiotti, L.~Knowles,
  A.~Estoup, M.~Panchal, J.~Corander, M.~Hickerson, et~al., 2010.
\newblock In defence of model-based inference in phylogeography.
\newblock Molecular Ecology 19:436--446.

\bibitem[{Beaumont et~al.(2002)Beaumont, Zhang, and Balding}]{beaumont02}
Beaumont, M.~A., W.~Zhang, and D.~J. Balding, 2002.
\newblock Approximate {B}ayesian computation in population genetics.
\newblock Genetics 162:2025--2035.

\bibitem[{Bertorelle et~al.(2010)Bertorelle, Benazzo, and Mona}]{bertorellel10}
Bertorelle, G., A.~Benazzo, and S.~Mona, 2010.
\newblock {ABC} as a flexible framework to estimate demography over space and
  time: some cons, many pros.
\newblock Molecular Ecology 19:2609--2625.

\bibitem[{Blum and Fran\c{c}ois(2010)}]{blumfrancois10}
Blum, M. G.~B. and O.~Fran\c{c}ois, 2010.
\newblock Non-linear regression models for {A}pproximate {B}ayesian
  {C}omputation.
\newblock Statistics and Computing 20:63--73.

\bibitem[{Capblancq et~al.(2015)Capblancq, Despr{\'e}s, Rioux, and
  Mav{\'a}rez}]{capblancq15}
Capblancq, T., L.~Despr{\'e}s, D.~Rioux, and J.~Mav{\'a}rez, 2015.
\newblock Hybridization promotes speciation in coenonympha butterflies.
\newblock Molecular Ecology 24:6209--6222.

\bibitem[{Cornuet et~al.(2010)Cornuet, Ravigne, and Estoup}]{cornuetetal10}
Cornuet, J.-M., V.~Ravigne, and A.~Estoup, 2010.
\newblock Inference on population history and model checking using {DNA}
  sequence and microsatellite data with the software {DIYABC} (v1.0).
\newblock BMC Bioinformatics 11:401.

\bibitem[{Csill\'ery et~al.(2010)Csill\'ery, Blum, Gaggiotti, and Fran{\c
  c}ois}]{csilleryetal10}
Csill\'ery, K., M.~G.~B. Blum, O.~E. Gaggiotti, and O.~Fran{\c c}ois, 2010.
\newblock {Approximate Bayesian Computation in practice}.
\newblock Trends Ecol Evol 25:410--418.

\bibitem[{Csill{\'e}ry et~al.(2012)Csill{\'e}ry, Fran{\c{c}}ois, and
  Blum}]{csillery12}
Csill{\'e}ry, K., O.~Fran{\c{c}}ois, and M.~G.~B. Blum, 2012.
\newblock abc: an {R} package for approximate {B}ayesian computation (abc).
\newblock Methods in ecology and evolution 3:475--479.

\bibitem[{Dong et~al.(2014)Dong, Zou, Lei, Liang, Li, and Yang}]{dong14}
Dong, F., F.-S. Zou, F.-M. Lei, W.~Liang, S.-H. Li, and X.-J. Yang, 2014.
\newblock Testing hypotheses of mitochondrial gene-tree paraphyly: unravelling
  mitochondrial capture of the streak-breasted scimitar babbler (pomatorhinus
  ruficollis) by the taiwan scimitar babbler (pomatorhinus musicus).
\newblock Molecular ecology 23:5855--5867.

\bibitem[{Elith and Leathwick(2009)}]{elith09}
Elith, J. and J.~R. Leathwick, 2009.
\newblock Species distribution models: ecological explanation and prediction
  across space and time.
\newblock Annual Review of Ecology, Evolution, and Systematics 40:677.

\bibitem[{Excoffier et~al.(2013)Excoffier, Dupanloup, Huerta-S{\'a}nchez,
  Sousa, and Foll}]{excoffier13}
Excoffier, L., I.~Dupanloup, E.~Huerta-S{\'a}nchez, V.~C. Sousa, and M.~Foll,
  2013.
\newblock Robust demographic inference from genomic and snp data.
\newblock PLoS Genetics 9:e1003905.

\bibitem[{Fran{\c{c}}ois and Laval(2011)}]{francois11}
Fran{\c{c}}ois, O. and G.~Laval, 2011.
\newblock Deviance information criteria for model selection in approximate
  {B}ayesian computation.
\newblock Statistical Applications in Genetics and Molecular Biology 10:1--25.

\bibitem[{Gelman et~al.(2014)Gelman, Carlin, Stern, and Rubin}]{gelman14}
Gelman, A., J.~B. Carlin, H.~S. Stern, and D.~B. Rubin, 2014.
\newblock Bayesian data analysis, vol.~2.
\newblock Taylor \& Francis.

\bibitem[{Gelman et~al.(1996)Gelman, Meng, and Stern}]{hill96}
Gelman, A., X.-L. Meng, and H.~Stern, 1996.
\newblock Discussion of `posterior predictive assessment of model fitness via
  realized discrepancies'.
\newblock Statistica sinica 6:767--773.

\bibitem[{Gruenstaeudl et~al.(2015)Gruenstaeudl, Reid, Wheeler, and
  Carstens}]{gruenstaeudl15}
Gruenstaeudl, M., N.~M. Reid, G.~L. Wheeler, and B.~C. Carstens, 2015.
\newblock Posterior predictive checks of coalescent models: P2c2m, an r
  package.
\newblock Molecular Ecology Resources .

\bibitem[{Hartig et~al.(2014)Hartig, Dislich, Wiegand, and Huth}]{hartig14}
Hartig, F., C.~Dislich, T.~Wiegand, and A.~Huth, 2014.
\newblock Technical note: Approximate {B}ayesian parameterization of a
  process-based tropical forest model.
\newblock Biogeosciences 11.

\bibitem[{Hickerson(2014)}]{hickerson14}
Hickerson, M.~J., 2014.
\newblock All models are wrong.
\newblock Molecular ecology 23:2887--2889.

\bibitem[{Hijmans(2012)}]{hijmans12}
Hijmans, R.~J., 2012.
\newblock Cross-validation of species distribution models: removing spatial
  sorting bias and calibration with a null model.
\newblock Ecology 93:679--688.

\bibitem[{Hudson(2002)}]{hudson02}
Hudson, R.~R., 2002.
\newblock {Generating samples under a Wright-Fisher neutral model of genetic
  variation}.
\newblock Bioinformatics 18:337--338.

\bibitem[{Johnson(2004)}]{johnson04}
Johnson, V.~E., 2004.
\newblock A bayesian chi2 test for goodness-of-fit.
\newblock Annals of Statistics Pp. 2361--2384.

\bibitem[{Kodandaramaiah and Wahlberg(2009)}]{kodandara09}
Kodandaramaiah, U. and N.~Wahlberg, 2009.
\newblock Phylogeny and biogeography of coenonympha butterflies (nymphalidae:
  Satyrinae)--patterns of colonization in the holarctic.
\newblock Systematic Entomology 34:315--323.

\bibitem[{Lagarrigues et~al.(2015)Lagarrigues, Jabot, Lafond, and
  Courbaud}]{lagarrigues15}
Lagarrigues, G., F.~Jabot, V.~Lafond, and B.~Courbaud, 2015.
\newblock Approximate {B}ayesian computation to recalibrate individual-based
  models with population data: Illustration with a forest simulation model.
\newblock Ecological Modelling 306:278--286.

\bibitem[{Laval et~al.(2010)Laval, Patin, Barreiro, and
  Quintana-Murci}]{laval10}
Laval, G., E.~Patin, L.~B. Barreiro, and L.~Quintana-Murci, 2010.
\newblock Formulating a historical and demographic model of recent human
  evolution based on resequencing data from noncoding regions.
\newblock PloS one 5:e10284.

\bibitem[{Lehmann(1993)}]{lehmann93}
Lehmann, E.~L., 1993.
\newblock The fisher, neyman-pearson theories of testing hypotheses: One theory
  or two?
\newblock Journal of the American Statistical Association 88:1242--1249.

\bibitem[{Li et~al.(2014)Li, Schlebusch, and Jakobsson}]{li14}
Li, S., C.~Schlebusch, and M.~Jakobsson, 2014.
\newblock Genetic variation reveals large-scale population expansion and
  migration during the expansion of bantu-speaking peoples.
\newblock Proceedings of the Royal Society of London B: Biological Sciences
  281:20141448.

\bibitem[{Meng(1994)}]{meng94}
Meng, X.-L., 1994.
\newblock Posterior predictive p-values.
\newblock The Annals of Statistics Pp. 1142--1160.

\bibitem[{Morales et~al.(2015)Morales, Mermoz, Gowda, and
  Kitzberger}]{morales15}
Morales, J.~M., M.~Mermoz, J.~H. Gowda, and T.~Kitzberger, 2015.
\newblock A stochastic fire spread model for north patagonia based on fire
  occurrence maps.
\newblock Ecological Modelling 300:73--80.

\bibitem[{Pelletier and Carstens(2014)}]{pelletier14}
Pelletier, T.~A. and B.~C. Carstens, 2014.
\newblock Model choice for phylogeographic inference using a large set of
  models.
\newblock Molecular ecology 23:3028--3043.

\bibitem[{Peter et~al.(2012)Peter, Huerta-Sanchez, and Nielsen}]{peter12}
Peter, B.~M., E.~Huerta-Sanchez, and R.~Nielsen, 2012.
\newblock Distinguishing between selective sweeps from standing variation and
  from a de novo mutation .

\bibitem[{Pritchard et~al.(1999)Pritchard, Seielstad, Perez-Lezaun, and
  Feldman}]{pritchard99}
Pritchard, J.~K., M.~T. Seielstad, A.~Perez-Lezaun, and M.~W. Feldman, 1999.
\newblock Population growth of human {Y} chromosomes: a study of {Y} chromosome
  microsatellites.
\newblock Molecular Biology and Evolution 16:1791--1798.

\bibitem[{Robins et~al.(2000)Robins, van~der Vaart, and Ventura}]{robins00}
Robins, J.~M., A.~van~der Vaart, and V.~Ventura, 2000.
\newblock Asymptotic distribution of p values in composite null models.
\newblock Journal of the American Statistical Association 95:1143--1156.

\bibitem[{Robinson et~al.(2014)Robinson, Bunnefeld, Hearn, Stone, and
  Hickerson}]{robinson14}
Robinson, J.~D., L.~Bunnefeld, J.~Hearn, G.~N. Stone, and M.~J. Hickerson,
  2014.
\newblock Abc inference of multi-population divergence with admixture from
  unphased population genomic data.
\newblock Molecular ecology 23:4458--4471.

\bibitem[{Roux et~al.(2013)Roux, Tsagkogeorga, Bierne, and Galtier}]{roux13}
Roux, C., G.~Tsagkogeorga, N.~Bierne, and N.~Galtier, 2013.
\newblock Crossing the species barrier: genomic hotspots of introgression
  between two highly divergent {\it {c}iona intestinalis} species.
\newblock Molecular {B}iology and {E}volution 30:1574--1587.

\bibitem[{Rykiel(1996)}]{rykiel96}
Rykiel, E.~J., 1996.
\newblock Testing ecological models: the meaning of validation.
\newblock Ecological modelling 90:229--244.

\bibitem[{Sj{\"o}din et~al.(2013)Sj{\"o}din, Sj{\"o}strand, Jakobsson, and
  Blum}]{sjodin12}
Sj{\"o}din, P., A.~E. Sj{\"o}strand, M.~Jakobsson, and M.~G.~B. Blum, 2013.
\newblock Resequencing data provide no evidence for a human bottleneck in
  {A}frica during the penultimate glacial period.
\newblock Molecular {B}iology and {E}volution 30:513--525.

\bibitem[{Voight et~al.(2005)Voight, Adams, Frisse, Qian, Hudson, and {Di
  Rienzo}}]{voight05}
Voight, B.~F., A.~M. Adams, L.~A. Frisse, Y.~Qian, R.~R. Hudson, and A.~{Di
  Rienzo}, 2005.
\newblock Interrogating multiple aspects of variation in a full resequencing
  data set to infer human population size changes.
\newblock Proc Natl Acad Sci USA 102:18508--18513.

\end{thebibliography}

\clearpage


\clearpage

\clearpage

\clearpage

%

%

\begin{center}
\begin{table}[ht]
{\small \begin{tabular}{c|cc|cc|cc}
~& \multicolumn{6}{c}{Null hypothesis}\\
  Truth & Bott. (SFS) & Exp. (SFS) & Bott (3 stat.)& Exp. (3 stat.)& 1 admix. & 2 admix.\\
%
%
%
  \hline
  Bott. (SFS) &  &  $21.5\%$ &&&&\\
  Exp. (SFS) & $53\%$ &  &&&&\\
  \hline
  Bott. (3 stat.)& &&  & $18\%$ ($17\%$) &&\\
  Exp. (3 stat.)& && $67\%$ ($72\%$) &  &&\\
  \hline
  1 admix. & && &&   & $19.5\%$\\
  2 admix. & && &&  $99\%$ & \\
  
  \end{tabular}}
\caption{Statistical power of the goodness-of-fit statistics $D_{\rm prior}$ for different problems of inference in evolutionary biology. The numbers given in parenthesis correspond to the statistical power obtained with $D_{\rm post}$. SFS stands for the site frequency spectrum, which is used as summary statistics and 3 stats. stands for the 3 statistics consisting of the average nucleotide diversity and the mean and variance (over loci) of Tajima's D.}
\label{tab:power}
\end{table}
\end{center}

  

\clearpage

\begin{center}
\begin{table}[ht]
\begin{tabular}{c|ccc}
~& Africa & Asia & Europe\\
\hline
const. &  0.21 (0.38)& 0.10 (0.07) & 0.02 (0.02)\\
bott. &  0.17 (0.43) & 0.86 (0.80) & 0.60 (0.50)\\
exp. & 0.55 (0.60)& 0.01 (0.02)& 0.00 (0.00)\\
 \end{tabular}
\caption{P-values obtained with the goodness-of-fit statistics $D_{\rm prior}$ and $D_{\rm post}$ (in parenthesis) for sequence data coming from 3 different populations \cite[]{voight05}. Data consist of the average nucleotide diversity and the mean and variance (over loci) of Tajima's D. Const. stands for constant population size, bott. stands for the bottleneck model, and exp. stands for the model of demographic expansion.}
\label{tab:data_gof}
\end{table}
\end{center}\end{document}


For the bottleneck simulations simulated with {\it fastsimcoal 2}, the population sizes before and after bottleneck were identical and drawn from a Log-Uniform(100,70000); the time (in generations BP) when the bottleneck started was drawn from a Uniform(10,5000); the duration of the bottleneck (in generations) from a Unifom(5,500); the coefficient of size reduction from Uniform(0.01,0.5). For the simulations of expansion, the population sizes before and after expansion were drawn from two Log-Uniform(100,70000) with the constraint of the present size being larger than the ancestral size; the time (in generations BP) when the expansion started was drawn from a Uniform(10,5000); the expansion was exponential and lasted until present.

For the simulations of admixture models, we assume the following prior distributions (times are given in number of generations that is assumed to be equal to the number of years). Population sizes for $P_1$ and $P_2$ are drawn from a Uniform($10^5$,$10^6$);  populations sizes for $P_3$ and $P_4$ are drawn from a Uniform($10^4$,$10^5$); the time of divergence between $P_1$ and $P_2$ is drawn from a Uniform($1.5\times 10^6$,$4\times 10^6$); the time of the most ancient admixture event is drawn from a Uniform($0$,$10^5$); the time of the most recent admixture event is drawn from a Uniform($0$,$50 \times 10^3$) with the constraint that it is smaller than the most ancient time; rates of admixture are  drawn from a Uniform($0.1$,$0.9$) distribution.